\documentclass[
submission
 , nomarks
]{dmtcs-episciences}

\usepackage[utf8]{inputenc}
\usepackage{subfigure}

\usepackage[round]{natbib}

\usepackage{stmaryrd}

\newcommand{\lambdaeab}{\lambda_{e}^{\rightarrow \Box}}
\newcommand{\lambdas}{\lambda_{\mathcal{S}}}
\newcommand{\minimlbox}{\mbox{Mini-ML}^{\Box}}

\newcommand{\kbox}{\mathop{\mathit{box}}}
\newcommand{\klet}{\mathop{\mathit{let}}}
\newcommand{\kletbox}{\mathop{\mathit{let\,box}}}
\newcommand{\kin}{\mathop{\mathit{in}}}
\newcommand{\unit}{\mathop{\mathit{unit}}}
\newcommand{\kunbox}{\mathop{\mathit{unbox}}}

\newcommand{\kif}{\mathop{\mathit{if}}}
\newcommand{\kthen}{\mathop{\mathit{then}}}
\newcommand{\kelse}{\mathop{\mathit{else}}}

\author{Martin Lester}
\title[Understanding the Expressive Power of Unhygienic Substitution via Combinatory Logic]{Understanding the Expressive Power of Unhygienic Substitution in Metaprogramming via Combinatory Logic}
\affiliation{
  Deparatment of Computer Science, University of Reading, United Kingdom}
\begin{document}
\publicationdetails{VOL}{2015}{ISS}{NUM}{SUBM}
\maketitle
\begin{abstract}
Recent work on combinatory logic demonstrates a compositional translation from
lambda calculus that gives meaning to open terms.
As the meaning of open terms is a key difficulty in the study of metaprogramming,
we investigate whether this idea can be extended to metaprogramming systems
with hygienic and unhygienic substitution,
specifically $\lambda_{e}^{\rightarrow \Box}$ and $\lambda_{\mathcal{S}}$.
We conclude that there is quantitative evidence to support the claim
that metaprogramming with unhygienic substitution
is more powerful than metaprogramming with hygienic substitution.
\end{abstract}

\section{Introduction}

Interest in combinatory logic as a practical formalism for implementing
functional programming languages \citep{DBLP:journals/spe/Turner79} dates back to Turner's bracket
abstraction algorithm \citep{DBLP:journals/jsyml/Turner79}.
Combinatory logic is appealing
because it removes the need for fiddly variable binding and substitution.

Although the system remains Turing-powerful, one might ask whether the
absence of variables somehow reduces the expressivity of the formalism.
Qualitatively, few would contest the claim that lambda calculus expressions
are easier to write than combinatory logic expressions. However, from a
quantitative perspective, the existence of near-linear translations from
lambda calculus into combinatory logic~\citep{DBLP:journals/ipl/Noshita85}
may suggest that the expressivity gap is not large.

The question remains whether this holds true for extensions of the lambda
calculus. Of particular interest to us are calculi that support some form of
metaprogramming.
Recently, \cite{DBLP:conf/flops/Kiselyov18} proposed an alternative, compositional translation
of lambda terms (when expressed using de Bruijn indices) into combinatory logic that is
linear under practical assumptions. Integral to the translation is a
semantics of lambda terms, \emph{even open lambda terms}, using terms of
combinatory logic.
This is significant as researchers working on semantics of metaprogramming
have sometimes struggled to give meaning to open lambda terms.
In this paper, we investigate whether this new translation
can be extended to metaprogramming systems,
and whether this reveals anything about their expressivity.

Several metaprogramming calculi have been proposed; one of the most
influential is $\lambdaeab$~\citep{DBLP:journals/jacm/DaviesP01}.
This calculus supports a form of extensional metaprogramming. Well-formed
code expressions can be composed to form new code expressions, which can
then be executed.
A type system that mirrors the structure of the modal
logic S4 ensures that code expressions formed in this way remain well-typed.
$\lambdaeab$ was the first typed metaprogramming calculus
to support execution of code \citep{DBLP:journals/corr/BergerT14a}.

Directly extending Kiselyov's translation to $\lambdaeab$
would not be immediately straightforward, as its modal type system
employs dual typing contexts \citep{DBLP:conf/lics/Kavvos17}.
However, there is a simple translation of the calculus into
unaugmented lambda calculus, essentially by treating code values as thunks.
One can compose this translation to unaugmented lambda calculus
with Kiselyov's translation to combinatory logic, effectively yielding a semantics for code terms
using combinatory logic.
As both translations are linear under reasonable assumptions, so too is their composition.

One restriction of $\lambdaeab$ is that unhygienic
substitution is not allowed: code expressions that are substituted cannot contain free
variables, and substitution cannot capture variables.
\cite{DBLP:conf/ecoop/BergerTU17} argue
that this is too restrictive for many practical metaprogramming
languages.
\cite{DBLP:conf/popl/KimYC06} present a calculus
$\lambda^{\mathit{sim}}_{\mathit{open}}$ with a modal type system that allows
unhygienic substitution;
\cite{DBLP:conf/popl/ChoiAYT11} simplify this to $\lambda_S$.

Could a similar approach work here?
No, as the calculus no longer enforces a fixed order of binding of variables, so the
semantics of an open code term would no longer be given by a single term of
combinatory logic, but by a set of $n!$ terms, corresponding to each possible
order of binding of the code's $n$ free variables. We argue that this
demonstrates that the expressivity gap between hygienic and unhygienic code
substitution is significant.

The remainder of the paper is structured as follows.
In Section~\ref{sec:bg}, we review combinatory logic,
translation of lambda calculus into combinatory logic,
and the metaprogramming systems $\lambdaeab$ and $\lambdas$.
Then, in Section~\ref{sec:translate}, we investigate the translation of these
systems into combinatory logic.
Next, in Section~\ref{sec:related}, we briefly summarise a few other results
on translation of lambda calculus into combinatory logic,
and on use of combinatory logic with metaprogramming.
Finally, in Section~\ref{sec:conc}, we summarise our results
and discuss their significance.

\section{Background}
\label{sec:bg}

We are concerned with the translation of systems of lambda calculus with metaprogramming features
into combinatory logic and the resulting increase in term size.
In this section, we review:
existing work on translation of lambda calculus into combinatory logic;
and two notable metaprogramming systems, $\lambdaeab$ and $\lambdas$,
whose translation into combinatory logic we will address in the following section.

\subsection{Review of Combinatory Logic}
\label{bg:combinatory}

Combinatory logic is a rewrite system over applicative terms, built from atoms called combinators.
Conventionally, the rewrites associated with each combinator correspond to closed terms of lambda calculus.
Here are the rewrite rules for some well-known combinators:

\[
\begin{array}{l}
S \> f \> g \> x \rightarrow f \> x \> (g \> x) \\
K \> x \> y \rightarrow x \\
I \> x \rightarrow x \\
B \> f \> g \> x \rightarrow f \> (g \> x) \\
C \> f \> g \> x \rightarrow f \> x \> g \\
\end{array}
\]

The \emph{bracket abstraction algorithm} defined by the following translation
$A_x[E]$ converts a lambda term $\lambda x . E$
into an extensionally equivalent term of combinatory logic:

\[
\begin{array}{l}
A_x [E_1 E_2] \mapsto S \> (A_x[E_1]) \> (A_x[E_2]) \\
A_x [x] \mapsto I \\
A_x [c] \mapsto K \> c \> \mbox{(where $c \neq x$)}
\end{array}
\]

As lambda calculus is a Turing-powerful system of computation, so too, then, is combinatory logic.
Furthermore, as the combinator $I$ is expressible as $S\>K\>K$,
just $S$ and $K$ is sufficient to achieve this.
In practice, $I$, $B$ and $C$ are often used, either for clarity,
or to avoid a cubic size increase in the translation from lambda calculus.
Intuitively, $B$, $C$ and $S$ serve as \emph{directors},
indicating whether an argument to a lambda expression should be passed
to the right half of its body, to the left half or to both halves.

Combinatory logic has attracted a great deal of interest because of its simplicity.
It avoids the most complicated part of lambda calculus, namely capture-avoiding substitution.
A line of research in the 1970s and 1980s investigated the use of combinatory logic
as the basis for hardware that could execute functional programs efficiently.
In this context, the size increase of translated terms is very important.

\subsection{Kiselyov's Compositional Translation}

A weakness of the standard bracket abstraction algorithm is that it is not compositional.
Consider a lambda term $e$, a context $C[\cdot]$
and the result of substituting $e$ into the hole in $C[\cdot]$, $C[e]$.
Now translate $e$ and $C[e]$ into equivalent combinatory logic terms $t_1$ and $t_2$,
using the bracket abstraction algorithm.
It is not necessarily the case that $t_1$ is a subterm of $t_2$, even though $e$ is a subterm of $C[e]$.

The reason for this is that the bracket abstraction transformation applies to terms with no nested lambdas.
It must be applied to each lambda in sequence, starting with the innermost.
The abstraction of outer lambdas modifies the terms generated during translation of inner lambdas.

Recent work by \cite{DBLP:conf/flops/Kiselyov18} addresses this weakness.
He presents a compositional bracket abstraction algorithm,
based on the structure of the type derivation of lambda terms represented using de Bruijn indices,
using a type system with a restricted form of weakening.

% show kiselyov's algorithm

The algorithm also works for untyped terms,
as it relies only on knowing the number of variables bound in the type environment.

A by-product of this system is a semantics for lambda terms, even open lambda terms, using combinatory logic.
The key idea is that the meaning of an open term is its closure under as many binding lambdas as necessary,
paired with the number of variables so bound.
Kiselyov calls this a \emph{denotational} semantics.
This point is debatable, as combinator terms are quite operational in flavour,
whereas the semantic domain in a denotational semantics is typically more abstract.
Nonetheless, the result is pertinent, as it is difficult to give meaning to open terms,
and these terms arise when dealing with metaprogramming systems.

\subsection{Davies and Pfenning's $\lambdaeab$ and $\minimlbox$}

One of the most influential works on metaprogramming is due to \cite{DBLP:journals/jacm/DaviesP01}.
They introduce metaprogramming systems which allow composition and execution of closed code.
In particular, they introduce a type system based on the modal logic S4,
which ensures that code with free variables cannot be executed.

They present two variants of lambda calculus.
The first, $\lambdaeab$, introduces two new primitives:
\begin{itemize}
\item $\kbox \> E$ --- treat expression $E$ as a code value;
\item $\kletbox \> u = E_1 \kin \> E_2$ --- unwrap code value $E_1$; bind to $u$ in $E_2$;
\end{itemize}
and a new reduction rule:
\[
\kletbox \> u = \kbox \> E_1 \kin \> E_2 \rightarrow [E_1/u]E_2
\]
For example, if $x$ and $y$ are two code values,
one can create a new code value that applies
the contents of $x$ to the contents of $y$ as follows:
\[
\kletbox \> u = x \kin \>
\kletbox \> v = y \kin \>
\kbox (u \> v)
\]
One can also write an \emph{eval} function that takes a code value as an argument and evaluates it:
\[
\lambda x . \kletbox \> u = x \kin \> u
\]

The second variant of lambda calculus is not named explicitly,
but forms the core of $\minimlbox$,
which replaces the relatively restricted $\kletbox$ with a new primitive $\kunbox$.
$\kunbox$ can occur inside $\kbox$ and functions similarly to Lisp's antiquotation.
When an $\kunbox$ expression evaluates to a code value,
it is spliced into the surrounding code value specified by the $\kbox$.
In their work, $\kunbox$ exists primarily for convenience, rather than to add expressive power.
Indeed, the semantics of $\kunbox$ is specified by translation into
$\kletbox$ within $\lambdaeab$.

\subsection{Unhygienic substitution in $\lambdas$}

% metaprogramming language
% type system

% metaprogramming language

% implicit vs explicit - what does it mean
% translation from explicit into pure lambda
% translation from implicit into explicit

\cite{DBLP:conf/popl/KimYC06} argue that $\minimlbox$ and many other metaprogramming systems
are too restrictive.
One of the weaknesses of $\minimlbox$ is that it does not allow
\emph{unhygienic} or \emph{capturing} substitution,
in which splicing a code value containing a free variable
into a context within a template where that variable is bound
results in a code value where the free occurrence becomes bound.
This feature is widely used in Lisp-like languages.
In response, they create $\lambda^{\mathit{sim}}_{\mathit{open}}$,
a system supporting both hygienic and unhygienic substitution.
They provide a type system that prevents code with free variables from being run.

In later work, \cite{DBLP:conf/popl/ChoiAYT11} simplify this to $\lambdas$,
which only supports unhygienic substitution.
The only unusual rule in its semantics is that for evaluation of $\kunbox$:
\[
\kunbox (\kbox v) \rightarrow v
\]
where $v$ is a value.
The subtlety is that free variables in $v$ may be captured by the surrounding context
during this step.
For example,
treating $\klet \> x = y \kin \> z$ as syntactic sugar for $(\lambda x . z) y$
with $x$ not free in $y$,
one can write:
\[
\klet a = \kbox \> y \kin
\kbox (\lambda x . \lambda y . (\kunbox \> a) x)
\]
which evaluates to $\kbox (\lambda x . \lambda y . y \> x)$, capturing the $y$
in $\kbox \> y$.

\citeauthor{DBLP:conf/popl/ChoiAYT11}'s work focuses on the translation of the metaprogramming calculus into
a lambda calculus augmented with records,
which they use as a means towards static analysis.
There are two aspects to this translation:
one is to make the use of metaprogramming explicit,
like \citeauthor{DBLP:journals/jacm/DaviesP01}'s translation from $\minimlbox$ to $\lambdaeab$;
the other is to use records to pass around and look up the values of captured variables.

\section{Translation}
\label{sec:translate}

We now consider whether it is possible to extend Kiselyov's translation
from lambda calculus into combinatory logic
to the metaprogramming systems outlined in the previous section,
and at what cost.

\subsection{Translating $\lambdaeab$}

As mentioned earlier, a common difficulty when dealing with metaprogramming systems
is giving meaning to open terms.
As Kiselyov's translation considers this,
we might reasonably hope that it can be extended to metaprogramming systems.
However, a possible difficulty soon becomes apparent.
Kiselyov's translation uses lambda terms expressed using de Bruijn indices,
with the indices controlling changes to the type environment.
In a metaprogramming system with open code values, this is no longer possible,
as the binding order of variables may be changed depending on the surrounding code
into which an open code value is spliced.

% give an example

However, in $\lambdaeab$, the type system ensures
that only closed code values may be spliced.
In fact, Davies and Pfenning propose the following mapping to translate
their language
into unaugmented lambda calculus:

\[
\begin{array}{lrclr}
\mbox{Types:} & \Box A & \mapsto & \unit \rightarrow A & \\
\mbox{Box:}   & \kbox \> E & \mapsto & \lambda x : \unit . E & \mbox{($x$ not free in $E$)}\\
\mbox{Let box:} & \kletbox \> E_1 \kin \> E_2 & \mapsto
  & (\lambda x : \unit \rightarrow . [x () / u] E_2 ) E_1 &
  \mbox{($x$ not free in $E_2$)}
\end{array}
\]

The translation, like Kiselyov's, is essentially compositional.
Composing two compositional translations gives a new compositional
translation, although there are two small points to consider.

Firstly, the substitution $[x () / u]$ in the case for $\kletbox$ is not quite compositional,
although it only affects leaves in the term tree for $E_2$.
This can easily be resolved in a combined translation,
as the typing derivation will make explicit that $u$ is a code variable and at what depth it is bound.

Secondly, the unit value $()$ and unit type $\unit$ are not usually present in combinatory logic.
However, as they serve no purpose other than to guard evaluation of code values, any combinator
(such as $I$) can be used in place of $()$.

The increase in term size because of the translation to unaugmented lambda calculus is linear,
at least when measuring term size as the number of nodes in the term tree.
When considering the unaugmented lambda calculus term expressed using de Bruijn indices,
a comparison is harder to make, as $\lambdaeab$ terms have not been expressed using de Bruijn indices.
In the case for $\kletbox$, if $u$ occurs deep within $E_2$, the de Bruijn index of $x$ may be large.
However, any comparable metric for $\lambdaeab$ terms will surely measure $u$ to be equally large,
as both $\kletbox \> u$ and $\lambda x : \unit$ are binders at the same height.
If we regard the translation as being linear in size then,
composed with Kiselyov's linear translation, so too is the combined translation.

The combination of these two translations is not particularly interesting
and does not provide any new insights.
The insight is that the restriction of code splicing to closed code values
makes the composed translation straightforward and, by any reasonable metric, linear.
Thus we argue that, quantitatively, the expressive power
added by metaprogramming in this case is minimal.

\subsection{Translating $\lambdas$}

In the presence of unhygienic substitution,
which allows open code to be spliced into surrounding code,
capturing free variables,
the problem becomes more complex.

We now need to reconsider what the meaning or denotation of an open term is.
For unaugmented lambda calculus, Kiselyov proposed that the denotation of an open term was
the same as the denotation of its closure under as many lambdas as necessary,
plus a record of how many variables has been bound.
For example,
in the context of $\lambda x . \lambda y . \lambda z . x \> y \> z$,
the denotation of $x \> y \> z$,
which we write $\llbracket x \> y \> z \rrbracket$, is
$\llbracket \lambda x . \lambda y . \lambda z . x \> y \> z \rrbracket$,
plus the fact that 3 variables are unbound.

This is fine for terms represented using de Bruijn indices,
as the order of binding of the variables is made explicit in the representation.
De Bruijn indices are sensible for a calculus that
respects the usual notion of $\alpha$-equivalence in lambda calculus,
but $\lambdas$ is not such a calculus.
In the presence of unhygienic substitution,
the order of binding of the variables depends on the context into which they are eventually substituted,
and there may be many contexts with different binding orders.

Consider, for example, the following code, where we have extended the
language with booleans and if/then/else, and $q_1$ and $q_2$ are boolean variables:
\[
\begin{array}{rcl}
\klet \> a & = & \kbox \> x \kin \\
\klet \> b & = & \kbox \> y \kin \\
\klet \> c & = & \kif \> q_1 \kthen a \kelse b \kin \\
\klet \> d & = & \kif \> q_2 \kthen \kbox \lambda x . \lambda y . x \> (\kunbox \> c) \\
      &   & \> \kelse \> \kbox \lambda y . \lambda x . x \> (\kunbox \> c)
\end{array}
\]
Here, $a$ and $b$ have different free variables.
In $c$, depending on the value of $q_1$, either $x$ or $y$ could be free.
Furthermore, whichever value $c$ takes, when the variable is captured in the construction of $d$,
the distance to the binding lambda depends dynamically on the value of $q_2$.

\subsubsection{Set of closures as denotation}

How can we resolve this problem?
One approach would be to view the denotation of an open term as being a \emph{set}
of denotations of closed terms:
one for each of the $n!$ possible binding orders of its $n$ free variables.
Thus, in the context of $\kbox \> x \> y \> z$, the denotation
$\llbracket x \> y \> z \rrbracket$ would be the set:
\[
\begin{array}{lcr}
\{ &
\llbracket \lambda x . \lambda y . \lambda z . x \> y \> z \rrbracket ,
\llbracket \lambda x . \lambda z . \lambda y . x \> y \> z \rrbracket ,
\llbracket \lambda y . \lambda x . \lambda z . x \> y \> z \rrbracket & , \\
& \llbracket \lambda y . \lambda z . \lambda x . x \> y \> z \rrbracket ,
\llbracket \lambda z . \lambda x . \lambda y . x \> y \> z \rrbracket ,
\llbracket \lambda z . \lambda y . \lambda x . x \> y \> z \rrbracket
& \}
\end{array}
\]
plus the information that 3 variables are unbound.
This is the crux of our argument that unhygienic substitution significantly increases expressive power.

\subsubsection{Canonical representative as denotation}

While this might be fine from a purely theoretical perspective,
the increase in size of the denotation makes it clearly unsuitable as the basis of a practical system.
There is also the problem of how to select the correct denotation from the set
at the point where an open code term is spliced.
Can we somehow recover a practically implementable system while following a similar approach?

One possibility
is to impose a total order on the variable names in the calculus.
The choice of order is arbitrary; we use alphabetical order of variable names.
In a computer implementation, one could use the quasi-lexicographic order
on the bit string encoding the variable name.
We can then define the denotation of an open term to be its closure under lambdas
that bind the free variables in decreasing order (plus the number of unbound variables).
For example, with $x < y < z$, the denotation $\llbracket x \> y \> z \rrbracket$
would be $\llbracket \lambda x . \lambda y . \lambda z . x \> y \> z \rrbracket$.

Regarding the set of denotations considered earlier as an equivalence class
$[ \llbracket \lambda x . \lambda y . \lambda z . x \> y \> z \rrbracket ]$,
we effectively choose to work with a canonical member of the class
$\llbracket \lambda x . \lambda y . \lambda z . x \> y \> z \rrbracket$ instead.
We can recover any other member of the class by applying a function
that permutes the canonical member's arguments.
For example,
$(\lambda f . \lambda z . \lambda y . \lambda x . f \> x \> y \> z)
\llbracket \lambda x . \lambda y . \lambda z . x \> y \> z \rrbracket =
\llbracket \lambda z . \lambda y . \lambda x . x \> y \> z \rrbracket$.

This still does not resolve the problem of how to choose
the correct denotation to use when splicing an open code term.
In some cases, the binding order imposed by splicing code might match the one we arbitrarily chose,
but in general it might not.
Indeed, as the same code could be spliced into two different contexts that impose different binding orders,
there is no possibility of always choosing the right order, even by design.

However, there may yet be hope for well-typed terms.
The type system of $\lambdas$ tracks which variables are free in open terms.
Thus, for a well-typed term, we know statically which variables are being captured by any splicing.
In \citeauthor{DBLP:conf/popl/ChoiAYT11}'s work,
where $\lambdas$ is translated into a calculus with records,
this is exploited to add bindings to a record that correspond to the values of captured variables.

If we simply wished to translate $\lambdas$ into unaugmented lambda calculus and thence to combinatory logic,
we could just follow \citeauthor{DBLP:conf/popl/ChoiAYT11}'s translation,
encode records in lambda calculus, then apply Kiselyov's translation.
However, in order to encode records, we would need to encode variable names to use as record field names,
as well as equality of names.
But this would just be reimplementing names.
What we want to do is capture the essence of the functionality that names provide
in this setting and no more.
This may be possible by adapting the translation to use our idea
for the denotation of code values with free variables,
permuting the arguments to code values when necessary.
However, the result would not clearly be compositional,
at least in the sense we defined earlier,
as at each step their translation of a term relies on manipulation of a stack of contexts,
not just the translations of its subterms.

\section{Related Work}
\label{sec:related}

The increase in size of a term,
when translated from lambda calculus to combinatory logic,
has been studied extensively.
Most recently, ~\cite{lachowski2018complexity} showed that,
in the worst case, the naive translation shown in Section~\ref{bg:combinatory}
suffers from a size increase of $\Theta(n^3)$.
Using $B$ and $C$, $\mathcal{O}(n \mathit{log} n)$ is achievable.
However, the increase in term size depends on the metric used for the size of the lambda term.
Most authors consider the size of a lambda term to be essentially the number of nodes in
its term tree, with all variables being of size 1.
Kiselyov's use of de Bruijn indices,
and his corresponding decision to treat $\mathbf{s}^n \mathbf{z}$ as being of size $n+1$,
inflates the size of the lambda term,
allowing him to claim an $\mathcal{O}(n)$ translation.

Metaprogramming has been studied extensively,
but rarely in the context of combinatory logic.
The SF-calculus ~\citep{DBLP:journals/jsyml/Given-WilsonJ11,DBLP:conf/icfp/JayP11}
extends combinatory logic with
a factorisation combinator $F$,
which can decompose an application into its two component terms.
(This is not a combinator in the traditional sense,
as it does not correspond to a closed term of lambda calculus.)
This imbues the calculus with a form of intensional metaprogramming.
A restriction on evaluation under $F$ ensures the system remains confluent.
The system allows a compositional translation of computable functions
into combinatory logic, even at higher order types.
In contrast, it is impossible to express certain functions,
such as \emph{parallel or}, in conventional lambda calculus or combinatory logic.

One further example of the study of metaprogramming and combinatory logic
concerns the study of the hardware design and theorem-proving language
ReFLect \citep{DBLP:journals/jfp/GrundyMO06}.
\cite{DBLP:journals/corr/MelhamCC13} show, in a combinatory setting,
that the ability to manipulate code values in the language
causes inconsistency in the type system,
unless certain restrictions are added.

\section{Conclusion}
\label{sec:conc}

We have shown that a compositional translation of lambda terms into combinatory logic
can easily be combined with a compositional translation from $\lambdaeab$ into lambda terms,
deriving a compositional translation from $\lambdaeab$ into combinatory logic.
The simplicity of this derivation, and the absence of any significant increase in term size,
lead us to argue that, quantitatively,
metaprogramming with hygienic code composition, as found in $\lambdaeab$, does not significantly increase
the expressive power of lambda calculus.

In contrast, applying this approach to $\lambdas$,
which features unhygienic substitution and does not support the usual notion of $\alpha$-equivalence,
is fraught with difficulty.
The obvious route towards adapting Kiselyov's translation to terms with free variables
leads to an $\mathcal{O}(n!)$ increase in the size of denotations.
Thus we argue that there is quantitative evidence to support the claim
that metaprogramming with unhygienic code composition, as found in $\lambdas$,
is significantly more powerful than with hygienic code composition.
There may be some hope of recovering a practical implementation of a metaprogramming system
in combinatory logic through a slightly different route, but it is unlikely to be compositional.

More generally, by investigating translations from systems with names into systems without names,
we have illuminated the differences between the ways in which names are used.
In the translation of unaugmented lambda calculus into combinatory logic,
we only really use the ability of names to be compared with other names for equality,
as exemplified by the use of directors in bracket abstraction algorithms.
In contrast, in attempting to translate a calculus with unhygienic substitution,
we also use the ability of names to be permuted.

%\nocite{*}
\bibliographystyle{abbrvnat}
% use the following instead if you encounter problems 
%\bibliographystyle{alpha}
\bibliography{paper}

\end{document}